\documentclass[11pt, oneside]{article}   	
\usepackage{geometry}                		
\usepackage[parfill]{parskip}    		
\usepackage{amssymb}
\usepackage{url}
\usepackage{natbib}
\usepackage[pdftex]{graphicx}
\title{Pseudo-$R^2$ statistics under complex sampling}
\author{Thomas Lumley}

\begin{document}
\maketitle

\begin{abstract}
Model summaries based on the ratio of fitted and null likelihoods have been proposed for generalised linear models, reducing to the familiar $R^2$ coefficient of determination in the Gaussian model with identity link.  In this note I show how to define the Cox--Snell and Nagelkerke summaries under arbitrary probability sampling designs, giving a design-consistent estimator of the population model summary.   I also show that for logistic regression models under case--control sampling the usual Cox--Snell and Nagelkerke $R^2$ are not design-consistent, but are systematically larger than would be obtained with a cross-sectional or cohort sample, even in settings where the weighted and unweighted logistic regression estimators are similar or identical. \\
\emph{Keywords: likelihood, logistic regression, case-control study, sampling weights}
\end{abstract}

\section{Background}
The coefficient of determination for a linear regression model is the proportional reduction in squared prediction error from using the model prediction instead of the mean. That is, if we have $n$ observations, a vector $X$ of $p$ predictors and a model 
$$E[Y|X=x]=\mu=\alpha+x\beta$$
we define
$$R^2=1-\frac{\sum_{i=1}^n (Y_i-\hat\mu_i)^2}{\sum_{i=1}^n (Y_i-\bar Y)^2}.$$
\citet[pp208--9]{cox-snell}, in an exercise, proposed a definition for binary $Y$ in terms of the likelihood ratio 
$$ R^{2}_{CS}=1-\left({L(0) \over L({\hat {\beta }})}\right)^{2/n}$$
where $L(0)$ is the maximised likelihood  for an intercept-only model 
$$E[Y]=\alpha$$
and $L(\hat {\beta })$ is the likelihood maximised over $\alpha$ and $\beta$. 

This definition reduces to the coefficient of determination in a linear-Normal model with $Y\sim N(\mu,\sigma^2)$ if the likelihood is also maximised over the variance parameter (so $-2\log L=2+n+n\log(2\pi\hat\sigma^2))$, though not if $L(0)/L(\hat\beta)$ is the likelihood ratio from a model with fixed $\sigma$ (as is common for generalised linear models).

\citet{Nagelkerke} noted that for Bernoulli data the maximum possible value of the likelihood is 1, and so the maximum possible value of $R^2_{CS}$ is $1-(L(0))^{{2/n}}$. He defined a rescaled pseudo-$R^2$
$$R^{2}=\left. R^2_{CS}\middle/ \left(1- L({0})\right)^{2/n}\right.$$
which attains the value 1 when prediction is perfect.  Nagelkerke's $R^2$ has become popular for logistic regression.

As maintainer of the {\sf survey} package for R\citep{package-survey}, I have been asked on more than one occasion how to compute Nagelkerke's $R^2$ under complex sampling. 
In this note I propose a definition in terms of a superpopulation parameter, and a design-based estimator. I show  the estimator consistently estimates the population value and reduces exactly to Nagelkerke's statistic under simple random sampling.  I also compare the proposed definition to the estimates obtained by naive use of the standard formula in unweighted logistic regression under case--control sampling, the standard analysis method in epidemiology and biostatistics.

Code and data for the examples is on \url{github.com/tslumley/pseudorsq}, and the estimators are available in the `survey' package starting from version 3.31--7.

\section{Definition}
Following \cite{lumley-scott-anzjs}, I will assume a finite population of size $N$ is an independent and identically distributed draw of vectors $(X, Y)$ from some (unknown) probability model with density $g(x,y)$, and a probability sample of size $n$ is taken from with known sampling probabilities $\pi_i$ for the $i$th individual.  I will write $E_p[\cdot]$ for expectations with respect to the superpopulation model and $E_\pi[\cdot]$ for expectations with respect to finite sampling, and write $w_i$ for the sampling weights $1/\pi_i$.

The analysis goal is to fit parametric regression models $f(y|x;\,\theta)$ for the marginal densities of $Y$ given $X$, where the parameter vector $\theta$ includes the regression intercept and slopes $(\alpha,\,\beta)$ and possibly other nuisance parameters.  I do not assume this parametric family necessarily contains the true model; the aim is to estimate the `least false' parameter $\theta^*$, the value minimising the Kullback--Leibler divergence between $f(y|x;\,\theta)$ and the true $g(\cdot)$.  Where there is a possibility of confusion, I will refer to $f(y|x;\,\theta)$ as the `working model' and the likelihood based on it as the `working likelihood'.  Let $\tilde\theta_N$ be the `census parameter', the value obtained by fitting the model $f(y|x;\,\theta)$ to the whole population.  By standard results on maximum likelihood estimation, $\tilde\theta_N\stackrel{p}{\to}\theta^*$. 

The parametric model is fitted to the complex sample by maximising the weighted loglikelihood 
$$\hat\ell(\theta) = \sum_{i\in\textrm{sample}} w_i \ell_i(\theta) = \sum_{i\in\textrm{sample}} w_i \log f(y_i|x_i;\,\theta).$$
The weighted likelihood is unbiased for the population working loglikelihood: 
$$E_\pi[\hat\ell(\theta)]=\sum_{i=1}^N \ell_i(\theta)$$
and its derivative is unbiased for the population working score function. 
Under mild conditions on the superpopulation distribution $g(\cdot)$ and the sampling design, the resulting estimator $\hat\theta$ is $\sqrt{n}$-consistent for $\tilde\theta_N$, and so also for $\theta^*$.   Because the main interest of the analyst is usually in the regression parameters $\beta$ I will abuse notation by writing $\hat\ell(\beta)$ for the value of $\hat\ell(\theta)$ with any other parameters estimated by maximum weighted likelihood.

We can now consider the Cox--Snell summary, $R^2_{CS}$. First, let $\ell_1(\beta)$ be the working loglikelihood for a single random observation from the superpopulation distribution, and define the superpopulation summary $\rho^2_{CS}$ by
\begin{equation}
\log\,(1- \rho^2_{CS}) =2E_p\left[\ell_1(\beta^*)-\ell_1(0)\right].
\label{true-rho}
\end{equation}
The ordinary Cox--Snell $R^2_{CS}$ for an iid sample is an estimator of $\rho^2_{CS}$ obtained by replacing the expectation with a population or sample average, and under iid sampling the consistency of $R^2_{CS}$ for $\rho^2_{CS}$ follows from the law of large numbers, the consistency of  maximum likelihood, and the smoothness of $\ell(\cdot)$.

The finite population has been constructed as an iid sample, so I can define the census (population) parameter $\tilde R^2_{CS}$ by
$$\log\,(1-\tilde R^2_{CS}) =\frac{2}{N}\left(\ell(\tilde\beta)-\ell(0)\right)=\frac{1}{N}\sum_{i=1}^N \left(\ell_i(\tilde\beta)-\ell_i(0)\right)$$
and obtain a design-based plug-in estimator
$$\log\,(1-\hat R^2_{CS}) =\frac{2}{\hat N}\left(\hat\ell(\hat\beta)-\hat\ell(0)\right)=\frac{\sum_{i\in{\textrm{sample}}} w_i \left(\ell_i(\hat\beta)-\ell_i(0)\right)}{\sum_{i\in{\textrm{sample}}} w_i }$$
That is, under complex sampling, the likelihood should be replaced by the weighted (pseudo)likelihood, and the $n$ in the Cox--Snell and Nagelkerke formulas should be replaced by the sum of the weights.   

The analysis so far has not assumed the model is correctly specified.  If the fitted parametric model family $f(\cdot;\,\theta)$ happened to include the true superpopulation distribution $g(\cdot)$, the right-hand side of equation~\ref{true-rho} would be the mutual information between $X$ and $Y$, or equivalently, the difference between the marginal and conditional entropy of $Y$.  This relationship to information theory gives another reason to regard $\rho^2_{CS}$ as a meaningful model summary.  

The design-based Cox--Snell statistic $\hat R^2_{CS}$ can now be rescaled to an upper limit of 1, obtaining a design-based version of the Nagelkerke pseudo-$R^2$.
$$\hat R^{2}=\left. \hat R^2_{CS}\middle/ \left(1- \exp(\hat \ell(0))\right)^{2/n}\right.$$

\section{Comparisons}
In this section I compare the proposed estimator either to a known or simulated population value or to the estimator obtained by ignoring the sampling. 

\subsection{A large multistage survey} 
NHANES, the National Health and Nutrition Examination Survey, is a series of large national surveys of the US civilian population conducted by the National Center for Health Statistics (NCHS). The survey has been run continuously since 1999, with data released for each two-year wave.  Each wave samples about 10,000 individuals, in a complex multistage, multiphase design, which is approximated for the purpose of public-use datasets by a two-stage sample.  In the approximate design, two city or county sampling units are taken from each of about 16 geographical strata, and a total of about 10,000 individual are sampled.  The survey oversamples people under 18 and over 60, and also oversamples racial/ethnic minority groups.  We will use data on blood pressure from the 2003--4 and 2005--6 waves of NHANES, in 18323 individuals for whom both dietary data and blood pressure were available\citep{nhanes0304,nhanes0506}.

Isolated Systolic Hypertension (ISH) is the form of hypertension most common in older people. It is defined by an elevated systolic blood pressure ($>140$ mmHg) with normal or low diastolic blood pressure ($<90$ mmHg), and indicates stiffening of walls of major arteries\citep{ish-nejm}. Following~\citet{lumley-scott-statsci} we fit a sequence of logistic regression models. The base model  used age as a linear spline with interior knots at 50 and 65 years. Successive models then added race/ethnicity as five categories, gender, a gender by age interaction, and reported dietary sodium intake.  The $R^2$ statistics for the models are shown in Table~\ref{ish-table}; for both the Cox--Snell and Nagelkerke statistics the design-based estimator has a slightly lower value than the estimator assuming simple random sampling.

\begin{table}[ht]
\caption{Standard and design-based Cox--Snell and Nagelkerke $R^2$ statistics in models for Isolated Systolic Hypertension, using data from NHANES 2003--2006. }
\centering
\begin{tabular}{lrr|rr}
 & $R^2_{CS}$ & $\hat R^2_{CS}$ & $R^2$ & $\hat R^2$ \\ 
  \hline
Age (linear spline, 3 knots) & 0.18 & 0.14 & 0.38 & 0.29 \\ 
  + race/ethnicity & 0.18 & 0.14 & 0.39 & 0.29 \\ 
  + gender& 0.18 & 0.14 & 0.39 & 0.30 \\ 
  + gender:age interaction & 0.18 & 0.15 & 0.40 & 0.31 \\ 
  + sodium intake & 0.18 & 0.15 & 0.40 & 0.31 \\ 
   \hline
\end{tabular}
\label{ish-table}
\end{table}

\subsection{Case--control sampling}
Logistic regression in case--control designs is probably the most frequently-used example of a generalised linear model under complex sampling.  Case--control sampling is used in the study of rare diseases: the small number of individuals with $Y=1$ are all sampled, but only a small fraction of those with $Y=0$. The standard analysis of a case--control sample in epidemiology is by unweighted logistic regression; the bias that would be expected from ignoring the sampling is confined to the intercept and the estimate of $\beta$ is the semiparametric MLE.  The unweighted analysis is sometimes, but not always, substantially more efficient than the weighted analysis. 

\subsubsection{Heuristics}
A heuristic analysis is useful.  Suppose that $P(Y=1)$ is very small in the population, and that $\pi_0\ll 1$ is chosen to give $m$ controls per case. The efficiency of this design (with unweighted analysis) relative to using the entire population is $m/(m+1)$, which can be chosen close to 1.   Almost all the (Fisher) information in the population is now present in the sample, so the likelihood ratio  between the null model and the model fitted by maximum likelihood will be approximately the same in the sample as in the population.   The ordinary Cox--Snell R-squared  in the population, $\tilde R^2_{CS}$, satisfies 
\begin{equation}
\log\,(1-\rho^2_{CS}) \approx\log\,(1-\tilde R^2_{CS}) =\frac{2}{N}\left(\ell(\tilde\beta)-\ell(0)\right)
\label{population}
\end{equation}
so in the unweighted model in the sample it will  satisfy
\begin{equation}
\log\,(1-R^2_{CS,\textrm{sample}}) \approx \frac{2}{n}\left(\ell(\tilde\beta)-\ell(0)\right)
\label{cc-sample}
\end{equation}
different by a factor of $N/n$.  That is, case--control sampling will dramatically deflate $\log(1-R^2_{CS})$ and so inflate $R^2_{CS}$ relative to prospective sampling from the same population.  In a less-ideal case--control scenario we would expect the sample likelihood ratio to be smaller than the population likelihood ratio --- but if it were smaller by  a factor of $n/N$, to cancel the bias, a case--control design would have no advantage over simple random sampling. 

The Nagelkerke correction will mitigate the sampling bias, because the rescaling factor is also affected by sampling bias and the biases partly cancel. However, 
 the sampling bias is multiplicative on the scale of the log likelihood ratio and the correction is multiplicative on the scale of $R^2$, a concave function of the log likelihood ratio. By Jensen's inequality, the ordinary Nagelkerke $R^2$ in the sample will still be biased upwards relative to the population value except when $R^2=0$ or $R^2=1$.  The design-based estimator $\hat R^2$ does not have these biases; it will estimate the same quantity under case--control sampling as under prospective sampling. 
 
\subsubsection{Simulation}
Table~\ref{sim-table} demonstrates this bias in a simulation. A population of size $10^5$ was simulated with a single $N(0,1)$ predictor, $X$, and a binary outcome $Y$ satisfying
$$\mathrm{logit}\,P(Y=1|X=x)=-6+x.$$
There were 389 cases in the simulated population, and control samples were drawn with 1, 2,5, 10, or 20 controls per case, giving control sampling fractions from 0.4\% to 7.8\%. As Table~\ref{sim-table} shows, the design-based estimators accurately reproduced the population $R^2$ statistics under all sampling fractions. The unweighted estimators displayed a substantial upwards bias that decreased with increasing sampling fraction, as expected from our heuristic argument. 

\begin{table}
\caption{Design-based and ordinary Cox--Snell and Nagelkerke $R^2$ in simulated case--control sampling}
\centering\begin{tabular}{rr|rrrr}
Matching & Sampling\\
 ratio & fraction &  $R^2_{CS}$ & $\hat R^2_{CS}$ & $R^2$ & $\hat R^2$\\ 
\hline
1 & 0.4\% & 0.21 & 0.0039 & 0.27 & 0.079\\
2 & 0.8\% & 0.19 & 0.0042 & 0.26 & 0.084\\
5 & 2.0\% & 0.11 & 0.0034 & 0.18 & 0.068\\
10 & 3.9\% &0.072 & 0.0036 & 0.16 & 0.072\\
20 & 7.8\% & 0.040 & 0.0036 & 0.13 & 0.072\\
\hline
\multicolumn{2}{l}{Population} & \multicolumn{2}{c}{0.0037} & \multicolumn{2}{c}{0.075}\\
\hline
\end{tabular}
\label{sim-table}
\end{table}

\subsubsection{A practical example}
To illustrate the practical relevance of this bias, consider the case--control study of oesphageal cancer in men from Ille-et-Vilaine, France, reported by \citet{tuyns} and later used by \citet{breslow-day} a study that is close to the heuristic ideal.  There are two publicly-available data sets: one with discrete variables for age (five groups),  tobacco consumption (four groups), and alcohol consumption (four groups), and the other with continuous variables for age and reported alcohol and tobacco consumption.   The study recruited between four and five controls per case; the control sampling fraction is not given explicitly but can be estimated from population data to be about 1/440. 

For both data sets, a model with main effects of each variable fits reasonably well. We consider this base model and a model adding a linear by linear interaction between  alcohol and tobacco consumption.  When using the grouped data, the estimated variances of $\hat\beta$ are very similar for the unweighted and design-weighted approaches; when using the continuous data, the unweighted estimate is substantially more precise.  

\begin{table}
\caption{Standard and design-based Cox--Snell and Nagelkerke $R^2$ statistics in models for oesphageal cancer risk based on age, alcohol consumption, and tobacco consumption, with data from a case--control sample. `Interaction' models include a linear alcohol by tobacco interaction term.}
\centering
\begin{tabular}{rrrrr}
 & $R^2_{CS}$ & $\hat R^2_{CS}$ & $R^2$ & $\hat R^2$\\ 
  \hline
Grouped data, main effects  & 0.14 & 0.0005 & 0.23 & 0.06 \\ 
Grouped data, interaction & 0.14 & 0.0005 & 0.23 & 0.06 \\ 
Continuous data, main effects & 0.26 & 0.0013 & 0.41 & 0.13 \\ 
Continuous data, interaction & 0.26 & 0.0013 & 0.41 & 0.14 \\ 
   \hline
\end{tabular}
\label{tuyns}
\end{table}

The $R^2$ estimates are shown in Table~\ref{tuyns}. As the heuristic analysis indicated, the ordinary pseudo-$R^2$ statistics are much larger than a design-based version, and so can be importantly biased relative to population or cohort statistics.  The bias exists both for the continuous-data models where the unweighted analysis is substantially more efficient, and the grouped-data models where there is little efficiency difference. 

\subsection{Informative two-phase sampling from a cohort}
\label{twophase}

Wilms' Tumour is a rare, largely treatable childhood cancer of the kidney, and the National Wilms' Tumor Study Group has run a series of clinical trials aiming to reduce the long-term side-effects of treatment while maintaining the cure rate\citep{nwts-1,nwts-2}. From a biostatistical viewpoint, one of the interesting aspects of Wilms' Tumour is that the histological classification (roughly, `cell abnormality') is difficult, and the study group central pathologist appears to be better at it than anyone else. That is, given the central lab histology, the local-hospital histology is not predictive of relapse, and the local-hospital histology can be treated as the central-lab histology plus random error.  There was obvious interest in studying Wilms' Tumour relapse with central-lab analysis of only a subset, rather than of all cases, and one of the data sets from the NWTG studies has become a standard example of two-phase sampling (eg, \citep{breslow-AJE-twophase,kulich-lin})

Table~\ref{nwtco}, compares the results for the full cohort to two  sampling schemes:  case--control sampling with equal numbers of cases and controsl, and balanced two-phase sampling with the same number sampled from all four cells of a $2\times 2$ local-histology by relapse table.  The design-based estimator is always close to the full cohort estimator; however, the unweighted estimator under case--control sampling is biased upwards.  The bias is less dramatic than in the previous example because the control sampling fraction is larger: about 10\% versus less than 1\%. 

\begin{table}
\caption{Models for relapse in Wilms' Tumor: using the full cohort, using a case-control sample, and using a two-phase sample based on outcome and predictors.}
\centering\begin{tabular}{lrrrr}
& $\hat R^2_{CS}$ & $R^2_{CS}$ & $\hat R^2$ & $R^2$\\
Case--control & 0.097 & 0.16 & 0.17 & 0.21\\\
Two-phase & 0.087 & --- & 0.16 & ---\\
\hline
Full cohort & \multicolumn{2}{c}{$0.086$} & \multicolumn{2}{c}{$0.16$}\\
\end{tabular}
\label{nwtco}
\end{table}

\section{Discussion}

There is some controversy about the usefulness of pseudo-$R^2$ measures for generalised linear models, but they are quite widely used.  In this note I have shown how to construct design-based versions of the Nagelkerke and Cox--Snell pseudo-$R^2$, which should be of use to survey statisticians.  I have also shown that the standard versions of these statistics when used for logistic regression on case--control samples do not estimate the same model summary as they would under prospective or cross-sectional sampling --- a fact that does not seem to be well known in biostatistics and epidemiology.  

\bibliographystyle{abbrvnat}
\bibliography{../Documents/TEX/survey,../Documents/TEX/raoscott}

\end{document}